\documentclass[twoside]{article}
\input ibvs3.sty

\usepackage{epsfig}
\usepackage{epstopdf}
\usepackage{longtable}
\usepackage{caption}

\newcommand{\kms}{km\,s$^{-1}$}
\newcommand{\etal}{$et~al.~$}

\begin{document}
\IBVSheadDOI{63}{6252}{21 Month 2018}

\IBVStitle{The Period Analysis of the Hierarchical System DI Peg}

\IBVSauth{OZUYAR, D.; ELMASLI, A.; CALISKAN, S.}

\IBVSinst{Ankara University, Faculty of Science, Dept. of Astronomy and Space Sciences, 06100, Tandogan, Ankara / Turkey, e-mail: dozuyar@ankara.edu.tr}

\SIMBADobjAlias{DI Peg}{HIP 116167}

\IBVSabs{The existence of an additional body around a binary system can be detected by the help of the light-travel time effect. Due to the motions of binary and the companion stars around the common mass center of the ternary system, the light-time effect produces an irregularity on the eclipse timings. Monitoring the variations in these timings, sub-stellar or planet companions orbiting around the binary system can be identified. In this paper, additional bodies orbiting the Algol type binary {DI Peg} are examined by using the archival eclipse timings including our CCD data observed at the Ankara University Kreiken Observatory. More than five hundred minimum times equivalent to around nine decades are employed to identify the orbital behavior of the binary system. The best fit to the timings shows that the orbital period of DI Peg has variation due to an integration of two sinusoids with the periods of $P_3 = 49.50\pm0.36$ yr and $P_4 = 27.40\pm0.24$ yr. The orbital change is thought to be most likely due to the existence of two M-type red dwarf companions with the masses of $M_3 =0.213 \pm 0.021$ M$_\odot$ and $M_4 = 0.151 \pm 0.008$ M$_\odot$, assuming that the orbits of additional bodies are co-planar with the orbit of the binary system. Also, the residuals of two sinusoidal fits still seem to show another modulation with the period of roughly $P = 19.5$ yr. The origin of this modulation is not clear and more observational data are required to reveal if the periodicity is caused by another object gravitationally bounded to the system.}

\begintext

\section{Introduction}
\label{intro}

Hierarchical multi-body star systems (Evans 1968) form from different ways, such as from interaction/capture in a globular star cluster (van den Berk \etal 2007), from a massive primordial disk involving accretion processes and/or local disk instabilities (Lim and Takakuwa 2006; Marzari \etal 2009) or from a non-hierarchical star system by angular momentum and energy exchange via mutual gravitational interactions (Reipurth 2000). These systems can be basically classified into two groups; circumbinary and circumstellar systems. In circumbinary systems, one or more additional bodies move around a binary star and they are known as companions on P-type orbits (Dvorak 1986). A transiting circumbinary planet, PH1b, around \hbox{{KIC 4862625}} which consists of two binary pairs; the quadruple systems \hbox{{HD 98800}} (Furlan \etal 2007) and {SZ Her} (Lee \etal 2012) can be given as  examples of such a hierarchy. On the other hand, the systems with companions orbiting one component of a binary pair are the other type of hierarchical systems (circumstellar or S-type configuration; Schwarz \etal 2011). The example of such a system can be found in Neuhäuser \etal (2007) and Chauvin \etal (2007).

A hierarchical circumbinary system can be detected by observing the timings of the mid-eclipse times of the binary companion.
The presence of an additional body causes a change in the relative distance of the eclipsing pair to the observer depending on the motion of the third body around the barycenter of the triple system. This binary wobble leads a periodic variation in conjunction times. As a result, the eclipses present lags or advances in the timings of minimum light (Irwin 1952). As known, the light-time effect is a geometrical feature and the third object produces a sinusoidal-like variation in the binary orbital. If the archival database is large and sufficient enough, this variation in eclipse timings provides an opportunity to understand the nature of the multi-body system (Pribulla \etal 2012). 

In this sense, space-based missions offer a unique opportunity for the discovery of companions orbiting eclipsing binaries. For example, Kepler provides continuous and highly homogeneous light curves over the time interval of four years. Thus, its  photometric observations enable new discoveries of multiple star systems, such as triple, quadruple or even quintuple ones. Indeed, there are a large number of multiple star systems identified from the Kepler observations. Conroy \etal (2014) present a catalog, which includes precise minimum times and third body signals for 1279 close binaries in the latest Kepler Eclipsing Binary Catalog. They find 236 binaries having third body signals. Borkovits \etal (2015) report {\sl O-C} analysis of 26 compact hierarchical triple stars in the Kepler field. Borkovits \etal (2016) identify the existence of a third body in 222 of 2600 Kepler binaries. The quadruple system KIC 7177553 (Lehmann \etal 2016) consists of two eccentric binaries with a separation of 0.4 arcsec (167 au). The outer orbit's period is in the range of 1000-3000 yr. Another quadruple star system EPIC 220204960 contains two slightly eccentric binaries with orbital periods of 13.27 and 14.41 days (Rappaport \etal 2017). These binaries are in a quadruple system with an outer period of 1 yr and a physical separation of ≤ 30 au. An example for a quintuple star system is EPIC 212651213 and EPIC 212651234 (Rappaport \etal 2016). In this system, EPIC 212651213 hosts two eclipsing binaries with orbital periods of 5.1 and 13.1 days. EPIC 212651234 is a single star with a projected physical separation of about 0.013 pc to EPIC 212651213. It is also stated that EPIC 212651213 and EPIC 212651234 are gravitationally bound to each other.

{DI Peg} ({HIP 116167}, {GSC 01175-00013}, {BD+14 5006}) was  discovered by Morgenroth (1934)  and identified to be an Algol type eclipsing binary (F4IV+ K4) by Rucinski (1967) and Lu (1992). From the photographic observations, Jensch (1934) determined the period of the system to be \hbox{$\sim 0_{\cdot}^\mathrm{d}711811$}. Rucinski (1967) analyzed the photoelectric observations of Kruszewski (1964) and derived the first orbital solutions. Based on the results, he suggested the existence of a third light which provided 24$\%$ contribution to the total light of the system. More photometric studies were performed by Chou and Kitamura (1968), Binnendijk (1973), Chaubey (1982), Lu (1992), and Yang \etal (2014).

Gaposchkin (1953) detected a variation in the orbital period of the star. Ahnert (1974) and Vinkó (1992) proposed a possible light-time effect in the system and they gave periods of $\sim62.4$ and $\sim22.1$ yr. By using the spectroscopic solutions, Lu (1992) determined the system parameters as $a = 4.14(0.05)$ R$_\odot$, $V_0 = +43.8(2.0)$ \kms, $K_1$ = 185.2(2.4) \kms, $K_2 = 109.0(2.1)$ \kms, $T_0$ = HJD 48213.8851(0.0022) and $q_\mathrm{sp}$ = 0.59(0.01). 

Rafert (1982) derived a downward quadratic ephemeris with a cyclic variation in the {\sl O-C} diagram. Unlike this, Hanna and Amin (2013) obtained a cyclic modulation with the period of 55 years, superimposed on an upward parabolic variation. The long-term orbital period increase was found to be $dP/dt = 0.17$ s/century and interpreted as a mass transfer from the evolved secondary component to the primary one with the rate of $1.52 \times 10^{-8}$ M$_\odot$/yr. The cyclic variation was attributed to a low-mass third body with the mass of $M_3 \sim 0.2200 \pm 0.0006$ M$_\odot$. The parameters of the third body were given as $e_3 = 0.77(7)$ and $w_3 = 300^{\circ} \pm  10^{\circ}$.
 
Recently, Yang \etal (2014) reproduced the photometric models by the help of new multi-color observations and previously published ones in literature. They determined the system parameters as $i = 89_{\cdot}^\mathrm{\circ}02\pm 0_{\cdot}^\mathrm{\circ}11$, $M_1 = 1.19(2)$ M$_\odot$, $M_2$ = 0.70(2) M$_\odot$, $L_1$ = 3.70(4) L$_\odot$, and $L_2$ = 0.53(2)~L$_\odot$. According to the results, they stated that the system had a low third light whose fill-out factor for the more massive component was $f_\mathrm{p} = 78.2(4)$. Their {\sl O-C} curve also indicated that the orbital period of DI\,Peg has changed in the complicated mode, such that the period of the star possibly showed two light-time orbits with the modulation periods of $P_3 \sim 54.6(5)$ yr and $P_4 \sim 23.0(6)$ yr, respectively. The masses of the inner and outer sub-stellar objects were given to be $M_{in}\sim 0.095$ M$_\odot$ and $M_{out}\sim 0.170$ M$_\odot$. On the basis of these results, Yang \etal (2014) suggested that the system has consisted of four objects.

The aim of this study is to perform a detailed period analysis of {DI Peg} for the parameter determination of the additional bodies in the system by using the new and all available archival minimum times. For this purpose, the paper is organized as follows; the observations are presented in Section~\ref{observations}, the analysis is described in Section~\ref{analysis}, the results related to the analysis are discussed in Section~\ref{conclusion}. 

\section{Observations}
\label{observations}

We observed {DI\,Peg} in V and R filters on the nights of 1 and 2 November 2017 at the Ankara University Kreiken Observatory. Observations were carried out by using an Apogee ALTA U47 + CCD camera  ($1024 \times 1024$ pixels) with Johnson UBVRI filters mounted on a 35 cm telescope. In the observing process, \hbox{{BD+14 5004}} was chosen as the comparison star (Table~1).  Bias, dark, and flat corrections were performed and all images were reduced by using the MaxIm DL software\footnote{https://diffractionlimited.com/help/maximdl/MaxIm-DL.htm}. The individual differential magnitudes were computed by subtracting the variable star from the comparison (V--C). The data covered two minima, the timings of which were determined as Min I = $2458060.4456 \pm 0.0001$ and Min II = $2458059.3779 \pm 0.0002$ with the method of Kwee and van Woerden (1956). The values were an average of the minimum times obtained in V and R colors during the same point.

\vskip 0.5cm

\centerline{Table 1. Spectral types, brightness, filters and exposure times are given for DI Peg and }
\centerline{its comparison star BD+14 5004.}
\vskip 3mm
\begin{center}
\begin{tabular}{lllccc}
\hline
Star&  &Spectral Type& V (mag) & Filters & Exposure Times (s) \\
\hline
DI Peg&Variable&F4-IV&9.52&R, V&35, 35\\
BD+14 5004&Comparison&K4&9.83&R, V&35, 35\\

\hline
\end{tabular}
\end{center}

\section{Analysis}
\label{analysis}

The {\sl O-C} diagram of {DI\,Peg} covering a time span of 88 years (Figure~1) was constructed from 85 primary, 14 secondary CCD; 45 primary, 9 secondary photoelectric; 17 primary photographic and 340 visual minimum times. These minima were collected from various observers listed in Table~2. The uncertainties of these values are not given in the table and can be accessed directly from their sources. The light elements of {DI\,Peg} were derived from the linear least-square fit applied to the CCD and photoelectric minimum times.

\begingroup
\tiny
\begin{longtable}{l}
\end{longtable}
  \endgroup

\begingroup
\tiny
\begin{longtable}{|l|l|l|l|l|l|l|l|}
\caption{All available minimum times of DI Peg in archives}
\label{mintable}\\
 &  &  &  &  &  &  &  \\
\endfirsthead

\multicolumn{8}{c}%
{{\bfseries}} \\
\endhead
Min. Time	&	Typ.	&	Meth.	&	Ref.			&	Min. Time	&	Typ.	&	Meth.	&	Ref.	\\
(HJD-2400000)	&	&		&			&	(HJD-2400000)	&		&		&		\\
\hline

25644.3150	&	1	&	pg	&	Guthnick \& Prager	;	AN 258	&	45201.4720	&	1	&	vi	&	 H.Peter 	;	 BBS 62 	\\
25918.3510	&	1	&	vi	&	 A.Jensch 	;	 AN 252.395 	&	45228.5220	&	1	&	vi	&	 N.Machkova 	;	 BRNO 26 	\\
26000.2330	&	1	&	vi	&	 A.Jensch 	;	 AN 252.395 	&	45231.3690	&	1	&	vi	&	 G.Mavrofridis 	;	 BBS 63 	\\
26249.3640	&	1	&	pg	&	 A.Jensch 	;	 AN 252.395 	&	45235.6450	&	1	&	vi	&	 G.Samolyk 	;	 AOEB 2 	\\
26266.4440	&	1	&	pg	&	 A.Jensch 	;	 AN 252.395 	&	45258.4170	&	1	&	vi	&	 H.Bohutinska 	;	 BRNO 26 	\\
26624.4580	&	1	&	pg	&	 A.Jensch 	;	 AN 252.395 	&	45554.5250	&	1	&	vi	&	 P.Svoboda 	;	 BRNO 26 	\\
26960.4600	&	1	&	vi	&	 A.Jensch 	;	 AN 252.395 	&	45579.4470	&	1	&	vi	&	 P.Svoboda 	;	 BRNO 26 	\\
26980.3840	&	1	&	vi	&	 A.Jensch 	;	 AN 252.395 	&	45609.3400	&	1	&	pg	&	 M.Dietrich 	;	 MVS 10.104 	\\
27738.4740	&	1	&	vi	&	 R.Szafraniec 	;	 AAC 4.81 	&	45609.3440	&	1	&	vi	&	 M.Zejda 	;	 BRNO 26 	\\
28432.4910	&	1	&	vi	&	 W.Opalski 	;	 BBG 1.47 	&	45624.2920	&	1	&	vi	&	 N.Stoikidis 	;	 BBS 69 	\\
28434.6270	&	1	&	vi	&	 W.Opalski 	;	 BBG 1.47 	&	45671.2750	&	1	&	vi	&	 P.Svoboda 	;	 BRNO 26 	\\
28452.4170	&	1	&	vi	&	 W.Opalski 	;	 BBG 1.47 	&	45915.4230	&	1	&	vi	&	 H.Peter 	;	 BBS 73 	\\
28454.5570	&	1	&	vi	&	 W.Opalski 	;	 BBG 1.47 	&	45976.6430	&	1	&	vi	&	 D.Williams 	;	 AOEB 2 	\\
28457.4050	&	1	&	vi	&	 W.Opalski 	;	 BBG 1.47 	&	45976.6500	&	1	&	vi	&	 S.Cook 	;	 AOEB 2 	\\
28459.5410	&	1	&	vi	&	 W.Opalski 	;	 BBG 1.47 	&	45981.6290	&	1	&	vi	&	 S.Cook 	;	 AOEB 2 	\\
28460.2510	&	1	&	vi	&	 W.Opalski 	;	 BBG 1.47 	&	45992.3030	&	1	&	vi	&	 A.Paschke 	;	 BBS 74 	\\
31273.3460	&	1	&	vi	&	 W.Zessewitsch 	;	 IODE 4.2.290 	&	46002.2610	&	1	&	vi	&	 A.Paschke 	;	 BBS 74 	\\
32441.4410	&	1	&	vi	&	 R.Szafraniec 	;	 AAC 4.81 	&	46019.3490	&	1	&	vi	&	 S.Krampol 	;	 BRNO 27 	\\
32794.4970	&	1	&	vi	&	 R.Szafraniec 	;	 AAC 4.113 	&	46028.6090	&	1	&	vi	&	 D.Williams 	;	 AOEB 2 	\\
32809.4430	&	1	&	vi	&	 R.Szafraniec 	;	 AAC 4.113 	&	46028.6110	&	1	&	vi	&	 G.Samolyk 	;	 AOEB 2 	\\
33170.3340	&	1	&	vi	&	 R.Szafraniec 	;	 AAC 5.5 	&	46029.3160	&	1	&	vi	&	 A.Paschke 	;	 BBS 74 	\\
33187.4120	&	1	&	vi	&	 R.Szafraniec 	;	 AAC 5.5 	&	46033.5850	&	1	&	vi	&	 S.Cook 	;	 AOEB 2 	\\
33538.3440	&	1	&	vi	&	 R.Szafraniec 	;	 AAC 5.7 	&	46038.5670	&	1	&	vi	&	 D.Williams 	;	 AOEB 2 	\\
33570.3780	&	1	&	vi	&	 R.Szafraniec 	;	 AAC 5.11 	&	46038.5680	&	1	&	vi	&	 G.Samolyk 	;	 AOEB 2 	\\
33871.4780	&	1	&	vi	&	 R.Szafraniec 	;	 AAC 5.11 	&	46043.5530	&	1	&	vi	&	 D.Williams 	;	 AOEB 2 	\\
33913.4740	&	1	&	vi	&	 A.Kruszewski 	;	 AA 6.140 	&	46290.5420	&	1	&	vi	&	 S.Stefanisko 	;	 BRNO 27 	\\
33916.3240	&	1	&	vi	&	 A.Kruszewski 	;	 AA 6.140 	&	46294.1170	&	1	&	vi	&	 T.Kato 	;	VSB 47 	\\
33918.4510	&	1	&	vi	&	 A.Kruszewski 	;	 AA 6.140 	&	46305.5010	&	1	&	vi	&	 A.Paschke 	;	 BBS 81 	\\
33928.4240	&	1	&	vi	&	 R.Szafraniec 	;	 AAC 5.11 	&	46320.4500	&	1	&	vi	&	 A.Paschke 	;	 BBS 81 	\\
34239.4900	&	1	&	vi	&	 R.Szafraniec 	;	 AAC 5.53 	&	46344.6500	&	1	&	vi	&	 S.Cook 	;	 AOEB 2 	\\
34254.4410	&	1	&	vi	&	 R.Szafraniec 	;	 AAC 5.191 	&	46350.3450	&	1	&	vi	&	 A.Paschke 	;	 BBS 81 	\\
34580.4550	&	1	&	vi	&	 R.Szafraniec 	;	 AAC 5.191 	&	46355.3240	&	1	&	vi	&	 O.Grugel 	;	BAVM 43 	\\
34664.4400	&	1	&	vi	&	 R.Szafraniec 	;	 AAC 5.191 	&	46360.3040	&	1	&	vi	&	 M.Dietrich 	;	 MVS 11.19 	\\
35010.3850	&	1	&	vi	&	 R.Szafraniec 	;	 AAC 5.194 	&	46360.3100	&	1	&	vi	&	 O.Grugel 	;	BAVM 43 	\\
35341.3830	&	1	&	vi	&	 R.Szafraniec 	;	 AA 6.143 	&	46382.3710	&	1	&	vi	&	 M.Dietrich 	;	 MVS 11.19 	\\
35366.3020	&	1	&	vi	&	 R.Szafraniec 	;	 AA 6.143 	&	46413.6980	&	1	&	vi	&	 G.Samolyk 	;	 AOEB 2 	\\
35699.4320	&	1	&	vi	&	 R.Szafraniec 	;	 AA 7.190 	&	46422.2380	&	1	&	vi	&	 A.Paschke 	;	 BBS 81 	\\
35719.3550	&	1	&	vi	&	 R.Szafraniec 	;	 AA 7.190 	&	46656.4230	&	1	&	vi	&	 M.Muller 	;	BAVM 46 	\\
35731.4490	&	1	&	vi	&	 R.Szafraniec 	;	 AA 7.190 	&	46678.4870	&	1	&	vi	&	 P.Hajek 	;	 BRNO 28 	\\
35746.4090	&	1	&	vi	&	 R.Szafraniec 	;	 AA 7.190 	&	46678.4900	&	1	&	vi	&	 A.Paschke 	;	 BBS 81 	\\
35838.2310	&	1	&	pg	&	 H.Huth 	;	 MVS 3.170 	&	46738.2760	&	1	&	vi	&	 D.Hanzl 	;	 BRNO 28 	\\
36079.5490 	&	1	&	pg	&	 H.Huth 	;	 MVS 3.170 	&	46743.2730	&	1	&	vi	&	 A.Stuhl 	;	 BRNO 31 	\\
36450.3900	&	1	&	vi	&	 R.Szafraniec 	;	 AA 9.49 	&	46759.6390	&	1	&	vi	&	 G.Samolyk 	;	 AOEB 2 	\\
36455.3779	&	1	&	vi	&	 J.Kordylewski 	;	 SAC 30.108 	&	46769.6070	&	1	&	vi	&	 G.Samolyk 	;	 AOEB 2 	\\
36462.4880 	&	1	&	pg	&	 H.Huth 	;	 MVS 3.170 	&	46774.5910	&	1	&	vi	&	 G.Samolyk 	;	 AOEB 2 	\\
36818.3880 	&	1	&	pg	&	 H.Huth 	;	 MVS 3.170 	&	46779.5640	&	1	&	vi	&	 G.Samolyk 	;	 AOEB 2 	\\
37193.5400	&	1	&	vi	&	 B.Czerlunczakiewic 	;	 AA 17.62 	&	46999.5200	&	1	&	vi	&	 G.Mavrofridis 	;	 BBS 86 	\\
37196.3810 	&	1	&	vis	&	 B.Czerlunczakiewic 	;	 EBC 1-32 	&	47014.4630	&	1	&	vi	&	 F.Hroch 	;	 BRNO 30 	\\
37196.3830 	&	1	&	vi	&	 J.Rodzinski 	;	 AA 18.332 	&	47014.4664	&	1	&	vi	&	 E.Wunder 	;	BAVM 50 	\\
37196.3910 	&	1	&	vis	&	 A.Slowik 	;	 EBC 1-32 	&	47029.4110	&	1	&	vi	&	 L.Prokesova 	;	 BRNO 30 	\\
37270.4040	&	1	&	vi	&	 F.Gerhart 	;	 AN 288.72 	&	47031.5490	&	1	&	vi	&	 J.Kolar 	;	 BRNO 30 	\\
37517.4080	&	1	&	vi	&	 A.Slowikowna 	;	 AA 17.62 	&	47034.4000	&	1	&	vi	&	 M.Jechumtal 	;	 BRNO 30 	\\
37522.3946	&	1	&	pe	&	 A.Kruszewski 	;	 AA 17.275 	&	47039.3790	&	1	&	vi	&	 O.Beck 	;	 BRNO 30 	\\
37523.4620	&	2	&	pe	&	 A.Kruszewski 	;	 AA 17.275 	&	47054.3330	&	1	&	vi	&	 G.Mavrofridis 	;	 BBS 86 	\\
37527.3776	&	1	&	pe	&	 A.Kruszewski 	;	 AA 17.275 	&	47066.4290	&	1	&	vi	&	 A.Paschke 	;	 BBS 86 	\\
37544.4610	&	1	&	pe	&	 A.Kruszewski 	;	 AA 17.275 	&	47091.3460	&	1	&	vi	&	 G.Mavrofridis 	;	 BBS 86 	\\
37556.5410 	&	1	&	vis	&	 H.Brancewicz 	;	 AA 17.62 	&	47107.7180	&	1	&	vi	&	 R.Hill 	;	 AOEB 2 	\\
37559.4096	&	1	&	pe	&	 A.Kruszewski 	;	 AA 17.275 	&	47387.4590	&	1	&	vi	&	 P.Adamek 	;	 BRNO 30 	\\
37626.3190	&	1	&	vi	&	 R.Gizinski 	;	BAVM 15 	&	47387.4610	&	1	&	vi	&	 A.Epple 	;	BAVM 52 	\\
37668.3160 	&	1	&	pg	&	 H.Huth 	;	 MVS 3.170 	&	47392.4390	&	1	&	vi	&	 P.Adamek 	;	 BRNO 30 	\\
37870.4760	&	1	&	vi	&	 H.Huth 	;	 MVS 3.170 	&	47464.3440	&	1	&	vi	&	 G.Samolyk 	;	 AOEB 2 	\\
37907.4920 	&	1	&	pg	&	 H.Huth 	;	 MVS 3.170 	&	47469.3150	&	1	&	vi	&	 G.Samolyk 	;	 AOEB 2 	\\
37932.3960	&	1	&	vi	&	 E.Pohl 	;	 AN 288.72 	&	47474.3180	&	1	&	vi	&	 H.Peter 	;	 BBS 90 	\\
37932.3970	&	1	&	vi	&	 F.Gerhart 	;	 AN 288.72 	&	47794.6200	&	1	&	vi	&	 G.Samolyk 	;	 AOEB 2 	\\
37932.4060 	&	1	&	pg	&	 H.Huth 	;	 MVS 3.170 	&	47851.5610	&	1	&	vi	&	 G.Samolyk 	;	 AOEB 2 	\\
37934.5370 	&	1	&	vi	&	 K.Klocke 	;	BAVM 15 	&	47853.6930	&	1	&	vi	&	 M.Smith 	;	 AOEB 2 	\\
37944.5060	&	1	&	vi	&	 J.Duball 	;	BAVM 15 	&	48123.4760	&	1	&	vi	&	 M.Copikova 	;	 BRNO 31 	\\
37947.3540	&	1	&	vi	&	 W.Braune 	;	BAVM 15 	&	48148.3950	&	1	&	vi	&	 J.Pietz 	;	BAVM 59 	\\
37956.6032	&	1	&	pe	&	Chou \& Kitamura	;	JKAS 1	&	48205.3360	&	1	&	vi	&	 J.Pietz 	;	BAVM 59 	\\
37983.6528	&	1	&	pe	&	Chou \& Kitamura	;	JKAS 1	&	48219.5690	&	1	&	vi	&	 G.Samolyk 	;	 AOEB 2 	\\
38253.4300	&	1	&	vi	&	 P.Flin 	;	 AA 17.62 	&	48266.5520	&	1	&	vi	&	 G.Samolyk 	;	 AOEB 2 	\\
38255.5610	&	1	&	vi	&	 H.Huth 	;	 MVS 3.170 	&	48480.8140	&	1	&	vi	&	 G.Samolyk 	;	 AOEB 2 	\\
38290.4530 	&	1	&	pg	&	 H.Huth 	;	 MVS 3.170 	&	48481.5240	&	1	&	vi	&	 J.Sojka 	;	 BRNO 31 	\\
38322.4780	&	1	&	vi	&	 V.Orlovius 	;	 AN 288.72 	&	48500.0280	&	1	&	vis	&	Hipparcos	;	ESA, 2001	\\
38399.3620	&	1	&	vi	&	 P.Hoffmann 	;	BAVM 18 	&	48506.4230	&	1	&	vi	&	 L.Honzik 	;	 BRNO 31 	\\
38591.5270 	&	1	&	pg	&	 H.Huth 	;	 MVS 3.170 	&	48543.8039	&	2	&	CCD	&	Hipparcos	;	ESA, 2001	\\
39006.5324	&	1	&	pe	&	 S.M.Rucinski 	;	 AA 17.275 	&	48545.5870	&	1	&	vi	&	 G.Samolyk 	;	 AOEB 2 	\\
39026.4630	&	1	&	vi	&	 W.Braune 	;	BAVM 18 	&	48554.8375	&	1	&	CCD	&	Hipparcos	;	ESA, 2001	\\
39046.3940	&	1	&	vi	&	 W.Braune 	;	BAVM 18 	&	48859.5040	&	1	&	vi	&	 J.Chlachula 	;	 BRNO 31 	\\
39056.3620	&	1	&	vi	&	 W.Braune 	;	BAVM 18 	&	48863.7660	&	1	&	vi	&	 D.Williams 	;	 AOEB 2 	\\
39061.3430	&	1	&	vi	&	 K.Locher 	;	 ORI 95 	&	48873.7330	&	1	&	vi	&	 R.Hill 	;	 AOEB 2 	\\
39352.4790	&	1	&	vi	&	 W.Braune 	;	BAVM 23 	&	48894.3760	&	1	&	vi	&	 R.Baule 	;	BAVM 62 	\\
39374.5440	&	1	&	vi	&	 K.Locher 	;	 ORI 100 	&	48935.3002	&	2	&	pe	&	 S.ozdemir 	;	IBVS 4380 	\\
39387.3600	&	1	&	vi	&	 W.Braune 	;	BAVM 23 	&	48939.2161	&	1	&	pe	&	 S.Selam 	;	IBVS 4380 	\\
39389.4960	&	1	&	vi	&	 W.Braune 	;	BAVM 23 	&	49215.4130	&	1	&	vi	&	 P.Stuchlik 	;	 BRNO 31 	\\
39407.2890	&	1	&	vi	&	 M.Seidl 	;	BAVM 23 	&	49224.6500	&	1	&	vi	&	 S.Cook 	;	 AOEB 2 	\\
39407.2930	&	1	&	vi	&	 K.Locher 	;	 ORI 100 	&	49241.7350	&	1	&	vi	&	 D.Williams 	;	 AOEB 2 	\\
39419.4010	&	1	&	vi	&	 S.Hazer 	;	 AN 291.113 	&	49246.3631	&	2	&	pe	&	 H.Ak 	;	IBVS 4380 	\\
39683.4680	&	1	&	vi	&	 K.Locher 	;	 ORI 103 	&	49248.4963	&	2	&	pe	&	 A.Akalin 	;	IBVS 4380 	\\
39827.2630	&	1	&	vi	&	 K.Locher 	;	 ORI 105 	&	49276.2546	&	2	&	pe	&	 H.Dundar 	;	IBVS 4380 	\\
40088.4990	&	1	&	vi	&	 F.Hromada 	;	 BRNO 9 	&	49277.3259	&	1	&	pe	&	 A.Akalin 	;	IBVS 4380 	\\
40114.8356	&	1	&	pe	&	 L.Binnendijk 	;	 AJ 78.97 	&	49333.5600	&	1	&	vi	&	 G.Samolyk 	;	 AOEB 2 	\\
40127.6488	&	1	&	pe	&	 L.Binnendijk 	;	 AJ 78.97 	&	49543.5440	&	1	&	vi	&	 C.Barani 	;	 BBS 108 	\\
40128.3600	&	1	&	vi	&	 P.Flin 	;	IBVS 328 	&	49543.5500 	&	1	&	vis	&	 F.Acerbi 	;	 BBS 107 	\\
40159.6796	&	1	&	pe	&	 L.Binnendijk 	;	 AJ 78.97 	&	49553.5085	&	1	&	pe	&	 B.Gurol 	;	IBVS 4380 	\\
40175.3430	&	1	&	vi	&	 W.Braune 	;	BAVM 23 	&	49602.6300	&	1	&	vi	&	 G.Samolyk 	;	 AOEB 2 	\\
40424.4746	&	1	&	pe	&	 N.Gudur 	;	IBVS 456 	&	49743.5640	&	1	&	vi	&	 G.Samolyk 	;	 AOEB 8 	\\
40471.4540	&	1	&	vi	&	 J.Silhan 	;	 BRNO 9 	&	49948.5775	&	1	&	vi	&	 M.Zibar 	;	 BRNO 32 	\\
40476.4370	&	1	&	vi	&	 J.Silhan 	;	 BRNO 9 	&	49950.7020	&	1	&	CCD	&	 S.Cook 	;	 AOEB 8 	\\
40483.5590	&	1	&	vi	&	 M.Fernandes 	;	BAVM 26 	&	50008.3599	&	1	&	CCD	&	 W.Kleikamp 	;	BAVM 90 	\\
40500.6394	&	1	&	pe	&	 L.Binnendijk 	;	 AJ 78.97 	&	50008.3603	&	1	&	CCD	&	 M.Wolf 	;	 BBS 110 	\\
40506.3380	&	1	&	vi	&	 K.Rausal 	;	 BRNO 12 	&	50013.3417	&	1	&	vi	&	 J.Cechal 	;	 BRNO 32 	\\
40512.7402	&	1	&	pe	&	 L.Binnendijk 	;	 AJ 78.97 	&	50044.6700	&	1	&	vi	&	 G.Samolyk 	;	 AOEB 8 	\\
40526.2640	&	1	&	vi	&	 K.Locher 	;	 ORI 116 	&	50050.3564	&	1	&	pe	&	 B.Gurol 	;	IBVS 4380 	\\
40725.5750	&	1	&	vi	&	 K.Locher 	;	 ORI 119 	&	50313.7370	&	1	&	vi	&	 G.Samolyk 	;	 AOEB 8 	\\
40772.5510	&	1	&	vi	&	 K.Locher 	;	 ORI 120 	&	50318.7140	&	1	&	CCD	&	 S.Cook 	;	 AOEB 8 	\\
40812.4130	&	1	&	vi	&	 W.Braune 	;	BAVM 25 	&	50368.5414	&	1	&	vi	&	 A.Dedoch 	;	 BRNO 32 	\\
40837.3269	&	1	&	pe	&	 O.Demircan 	;	IBVS 530 	&	50376.3686	&	1	&	CCD	&	 W.Kleikamp 	;	BAVM 102 	\\
40837.3290	&	1	&	vi	&	 W.Braune 	;	BAVM 25 	&	50396.3000	&	1	&	vi	&	 M.Dietrich 	;	BAVM 101 	\\
40837.3300	&	1	&	vi	&	 J.Hubscher 	;	BAVM 25 	&	50423.3560	&	1	&	vi	&	 D.Girrbach 	;	BAVM 101 	\\
40839.4630	&	1	&	vi	&	 R.Diethelm 	;	 ORI 121 	&	50667.4989	&	1	&	vi	&	 J.Polak 	;	 BRNO 32 	\\
40854.4130	&	1	&	vi	&	 M.Geseova 	;	 BRNO 12 	&	50672.4793	&	1	&	pe	&	 D.Husar 	;	BAVM 111 	\\
40856.5400	&	1	&	vi	&	 M.Geseova 	;	 BRNO 12 	&	50672.4805	&	1	&	pe	&	 W.Ogloza 	;	IBVS 4534 	\\
40859.3930	&	1	&	pe	&	 C.Endres 	;	IBVS 530 	&	50672.4909	&	1	&	vi	&	 J.Minar 	;	 BRNO 32 	\\
40859.3960	&	1	&	pg	&	 P.Ahnert 	;	 MVS 6.9 	&	50712.3428	&	1	&	pe	&	 D.Husar 	;	BAVM 111 	\\
40886.4480	&	1	&	vi	&	 H.Gese 	;	 BRNO 12 	&	50716.6150	&	1	&	vi	&	 G.Samolyk 	;	 AOEB 8 	\\
40911.3530	&	1	&	vi	&	 K.Locher 	;	 ORI 122 	&	50717.3278	&	1	&	vi	&	 L.Brat 	;	 BRNO 32 	\\
40921.3240	&	1	&	vi	&	 K.Locher 	;	 ORI 122 	&	50717.3305	&	1	&	vi	&	 P.Sobotka 	;	 BRNO 32 	\\
41155.5040	&	1	&	vi	&	 L.Frasinski 	;	IBVS 584 	&	50717.3370	&	1	&	pg	&	 M.Dietrich 	;	BAVM 113 	\\
41177.5740	&	1	&	vi	&	 K.Locher 	;	 ORI 126 	&	50719.4618	&	1	&	pe	&	 D.Husar 	;	BAVM 111 	\\
41210.3240	&	1	&	vi	&	 H.Peter 	;	 ORI 127 	&	50754.3480	&	1	&	vi	&	 R.Meyer 	;	BAVM 113 	\\
41232.3940	&	1	&	vi	&	 K.Locher 	;	 ORI 127 	&	51035.4000	&	1	&	pe	&	 B.Gurol 	;	IBVS 5069 	\\
41247.3320	&	1	&	vi	&	 K.Locher 	;	 ORI 129 	&	51045.4699	&	1	&	vi	&	 M.Vetrovcova 	;	 BRNO 32 	\\
41267.2632	&	1	&	vi	&	 W.Braune 	;	BAVM 25 	&	51076.7940	&	1	&	vi	&	 D.Williams 	;	 AOEB 8 	\\
41513.5560	&	1	&	vi	&	 K.Locher 	;	 BBS 4 	&	51079.6400	&	1	&	vi	&	 D.Williams 	;	 AOEB 8 	\\
41550.5620	&	1	&	vi	&	 K.Locher 	;	 BBS 5 	&	51084.6290	&	1	&	vi	&	 G.Samolyk 	;	 AOEB 8 	\\
41563.3810	&	1	&	vi	&	 H.Peter 	;	 BBS 5 	&	51141.5690	&	1	&	vi	&	 G.Samolyk 	;	 AOEB 8 	\\
41565.5120	&	1	&	vi	&	 K.Locher 	;	 BBS 5 	&	51422.0120	&	1	&	CCD	&	A.Paschke 	;	Amateur	\\
41580.4600	&	1	&	vi	&	 K.Locher 	;	 BBS 5 	&	51432.7010	&	1	&	vi	&	 D.Williams 	;	 AOEB 8 	\\
41595.4070	&	1	&	vi	&	 R.Diethelm 	;	 BBS 6 	&	51433.4096	&	1	&	CCD	&	 L.Kral 	;	 BRNO 32 	\\
41597.5432	&	1	&	vi	&	 W.Quester 	;	BAVM 26 	&	51452.6310	&	1	&	vi	&	 G.Samolyk 	;	 AOEB 8 	\\
41605.3720	&	1	&	vi	&	 W.Quester 	;	BAVM 26 	&	51467.5790	&	1	&	vi	&	 D.Williams 	;	 AOEB 8 	\\
41605.3730	&	1	&	vi	&	 K.Locher 	;	 BBS 6 	&	51807.4721	&	2	&	CCD	&	 W.Kleikamp 	;	BAVM 152 	\\
41605.3780	&	1	&	vi	&	 H.Peter 	;	 BBS 6 	&	51818.5020	&	1	&	CCD	&	 H.Achterberg 	;	BAVM 152 	\\
41657.3370	&	1	&	vi	&	 R.Diethelm 	;	 BBS 7 	&	51842.7060	&	1	&	vi	&	 R.Hill 	;	 AOEB 8 	\\
41682.2470	&	1	&	vi	&	 J.Hubscher 	;	BAVM 26 	&	51868.3321	&	1	&	CCD	&	 M.Dietrich 	;	BAVM 152 	\\
41682.2500	&	1	&	vi	&	 W.Braune 	;	BAVM 26 	&	52168.7180	&	1	&	vi	&	 D.Williams 	;	 AOEB 8 	\\
41921.4270	&	1	&	vi	&	 Z.Pokorny 	;	 BRNO 17 	&	52203.5970	&	1	&	vi	&	 D.Williams 	;	 AOEB 8 	\\
41928.5370	&	1	&	vi	&	 W.Quester 	;	BAVM 28 	&	52278.3363	&	1	&	CCD	&	 G.Maintz 	;	BAVM 152 	\\
41931.3750	&	1	&	vi	&	 R.Germann 	;	 BBS 11 	&	52530.3191	&	1	&	CCD	&	 M.Dietrich 	;	BAVM 158 	\\
41931.3930	&	1	&	vi	&	 I.Kohoutek 	;	 BRNO 17 	&	52542.7862	&	2	&	CCD	&	 Karska \& Maciejewski 	;	IBVS 5380 	\\
41941.3530	&	1	&	vi	&	 H.Peter 	;	 BBS 11 	&	52567.3312	&	1	&	CCD	&	 U.Schmidt 	;	BAVM 158 	\\
41983.3490	&	1	&	pg	&	 P.Ahnert 	;	 MVS 7.38 	&	52572.6843	&	2	&	CCD	&	 Karska \& Maciejewski 	;	IBVS 5380 	\\
41983.3560	&	1	&	vi	&	 J.Hudec 	;	 BRNO 17 	&	52573.0329	&	1	&	CCD	&	 Karska \& Maciejewski 	;	IBVS 5380 	\\
41988.3210	&	1	&	vi	&	 R.Germann 	;	 BBS 12 	&	52594.3820	&	1	&	pe	&	 T.Tanriverdi et al. 	;	IBVS 5407 	\\
42008.2630	&	1	&	vi	&	 H.Peter 	;	 BBS 12 	&	52843.5166	&	1	&	CCD	&	 B.Gurol et al. 	;	IBVS 5791 	\\
42274.4860	&	1	&	vi	&	 J.Hudec 	;	 BRNO 20 	&	52848.5024	&	1	&	vi	&	 L.Marcin 	;	OEJV 0074 	\\
42289.4270	&	1	&	vi	&	 H.Peter 	;	 BBS 17 	&	52848.5081	&	1	&	vi	&	 J.Pcola 	;	OEJV 0074 	\\
42289.4290	&	1	&	pe   	&	 O.Demircan 	;	IBVS 1053 	&	52888.3606	&	1	&	CCD	&	 T.Krajci 	;	IBVS 5592 	\\
42301.5400	&	1	&	vi	&	 J.Hudec 	;	 BRNO 20 	&	52903.3083	&	1	&	CCD	&	 M.Dietrich 	;	BAVM 172 	\\
42304.3760	&	1	&	vi	&	 R.Germann 	;	 BBS 17 	&	52908.2924	&	1	&	CCD	&	 M.Dietrich 	;	BAVM 172 	\\
42304.3960	&	1	&	vi	&	 M.Vlcek 	;	 BRNO 20 	&	52911.1395	&	1	&	CCD	&	 Nakajima 	;	VSB 42 	\\
42403.3170	&	1	&	vi	&	 K.Locher 	;	 BBS 19 	&	52950.2871	&	1	&	CCD	&	 B.Schlereth 	;	BAVM 172 	\\
42403.3220	&	1	&	vi	&	 H.Peter 	;	 BBS 19 	&	52986.5913	&	1	&	CCD	&	 S.Dvorak 	;	 AOEB 11 	\\
42403.3240	&	1	&	vi	&	 R.Diethelm 	;	 BBS 19 	&	52993.7110	&	1	&	vi	&	 G.Samolyk 	;	 AOEB 11 	\\
42739.2950	&	1	&	vi	&	 W.Braune 	;	BAVM 29 	&	53001.5420	&	1	&	vi	&	 D.Williams 	;	 AOEB 11 	\\
42739.3000	&	1	&	vi	&	 H.Peter 	;	 BBS 24 	&	53236.4399	&	1	&	vis	&	 J.Cernu 	;	OEJV 0074 	\\
42754.2470	&	1	&	vi	&	 H.Peter 	;	 BBS 25 	&	53236.4400	&	1	&	pe	&	 B.Albayrak et al. 	;	IBVS 5649 	\\
42776.2960	&	1	&	vi	&	 R.Germann 	;	 BBS 25 	&	53236.4476	&	1	&	vi	&	 M.Zdvoruk 	;	OEJV 0074 	\\
42786.2710	&	1	&	vi	&	 R.Germann 	;	 BBS 26 	&	53251.3810	&	1	&	vi	&	 R.Obertrifter 	;	BAVM 202 	\\
42786.2750	&	1	&	vi	&	 H.Peter 	;	 BBS 26 	&	53251.3840	&	1	&	vi	&	 G.-U.Flechsig 	;	BAVM 174 	\\
42796.2400	&	1	&	vi	&	 H.Peter 	;	 BBS 26 	&	53251.3860	&	1	&	vi	&	 K.Rutz 	;	BAVM 174 	\\
42990.5700	&	1	&	vi	&	 K.Locher 	;	 BBS 29 	&	53251.3910	&	1	&	vi	&	 W.Braune 	;	BAVM 174 	\\
42993.4120	&	1	&	vi	&	 K.Locher 	;	 BBS 29 	&	53262.4225	&	2	&	CCD	&	 F.Agerer 	;	BAVM 173 	\\
43013.3510	&	1	&	vi	&	 R.Germann 	;	 BBS 29 	&	53265.6239	&	1	&	vi	&	 W.Ogloza et al. 	;	IBVS 5843 	\\
43015.4802	&	1	&	pe	&	 J.Ebersberger 	;	IBVS 1358 	&	53267.7510	&	1	&	CCD	&	 W.Ogloza et al. 	;	IBVS 5843 	\\
43015.4840	&	1	&	vi	&	 P.Simecek 	;	 BRNO 21 	&	53267.7592	&	1	&	CCD	&	 G.Samolyk 	;	 AOEB 11 	\\
43034.7010	&	1	&	vi	&	 G.Samolyk 	;	 AOEB 2 	&	53272.7416	&	1	&	CCD	&	 W.Ogloza et al. 	;	IBVS 5843 	\\
43040.3980	&	1	&	vi	&	 K.Locher 	;	 BBS 30 	&	53282.7068	&	1	&	CCD	&	 W.Ogloza et al. 	;	IBVS 5843 	\\
43069.5700	&	1	&	vi	&	 E.Halbach 	;	 AOEB 2 	&	53285.5570	&	1	&	vi	&	 G.Chaple 	;	 AOEB 11 	\\
43069.5830	&	1	&	vi	&	 G.Samolyk 	;	 AOEB 2 	&	53290.5400	&	1	&	vi	&	 G.Chaple 	;	 AOEB 11 	\\
43071.0029	&	1	&	pe	&	 H.D.Kennedy 	;	IBVS 2118 	&	53292.6790	&	1	&	vi	&	 C.Stephan 	;	 AOEB 11 	\\
43112.2910	&	1	&	vi	&	 R.Germann 	;	 BBS 31 	&	53317.5870 	&	1	&	pe	&	 G.Lubcke 	;	 JAAVSO 41;328 	\\
43134.3600	&	1	&	vi	&	 R.Germann 	;	 BBS 31 	&	53325.4174	&	1	&	CCD	&	 W.Quester 	;	BAVM 173 	\\
43154.2880	&	1	&	vi	&	 R.Germann 	;	 BBS 32 	&	53614.4169	&	1	&	CCD	&	 V.Bakis et al. 	;	IBVS 5662 	\\
43311.5940	&	1	&	vi	&	 K.Locher 	;	 BBS 33 	&	53619.3969	&	1	&	vi	&	 P.Hejduk 	;	OEJV 0074 	\\
43341.4850	&	1	&	vi	&	 K.Vojtek 	;	 BRNO 21 	&	53634.3450	&	1	&	CCD	&	 M.Dietrich 	;	BAVM 178 	\\
43371.3870	&	1	&	vi	&	 R.Germann 	;	 BBS 34 	&	53645.0238	&	1	&	CCD	&	 Kubotera 	;	VSB 44 	\\
43391.3190	&	1	&	vi	&	 R.Germann 	;	 BBS 35 	&	53645.7354	&	1	&	CCD	&	 G.Samolyk 	;	 AOEB 11 	\\
43393.4570	&	1	&	vi	&	 K.Locher 	;	 BBS 35 	&	53671.3609	&	1	&	CCD	&	 R.Ehrenberger 	;	OEJV 0074 	\\
43393.4730	&	1	&	vi	&	 P.Ivan 	;	 BRNO 21 	&	53674.9210	&	1	&	vi	&	 Hirosawa 	;	VSB 44 	\\
43403.4350	&	1	&	vi	&	 P.Ivan 	;	 BRNO 21 	&	53728.3061	&	1	&	CCD	&	 J.Coloma 	;	 AOEB 11 	\\
43425.4940	&	1	&	vi	&	 K.Vojtek 	;	 BRNO 21 	&	53945.4760 	&	1	&	CCD	&	 K.Rutz 	;	BAVM 187 	\\
43433.3230	&	1	&	vi	&	 D.Lichtenknecker 	;	BAVM 31 	&	53967.4772	&	2	&	CCD	&	 S.Parimucha et al. 	;	IBVS 5777 	\\
43434.0295	&	1	&	pe	&	 H.D.Kennedy 	;	IBVS 2118 	&	53991.3226	&	1	&	vi	&	 S.Dogru et al. 	;	IBVS 5746 	\\
43435.4610	&	1	&	vi	&	 D.Lichtenknecker 	;	BAVM 31 	&	53992.3940	&	1	&	vi	&	 W.Braune 	;	BAVM 187 	\\
43455.3900	&	1	&	vi	&	 J.Soukopova 	;	 BRNO 21 	&	53993.1031	&	1	&	CCD	&	 K.Nagai et al. 	;	VSB 45 	\\
43460.3740	&	1	&	vi	&	 D.Sasselov 	;	 BRNO 21 	&	54016.5920	&	1	&	vi	&	 G.Chaple 	;	 AOEB 12 	\\
43490.2640	&	1	&	vi	&	 J.Mrazek 	;	 BRNO 21 	&	54023.7150	&	1	&	vi	&	 D.Williams 	;	 AOEB 12 	\\
43495.2440	&	1	&	vi	&	 R.Germann 	;	 BBS 36 	&	54024.4239	&	1	&	CCD	&	 F.Agerer 	;	BAVM 183 	\\
43517.3180	&	1	&	vi	&	 R.Germann 	;	 BBS 36 	&	54027.2706	&	1	&	CCD	&	 R.Ehrenberger 	;	OEJV 0074 	\\
43689.5710	&	1	&	vi	&	 K.Locher 	;	 BBS 37 	&	54032.9670	&	1	&	vi	&	 K.Nagai et al. 	;	VSB 45 	\\
43724.4540	&	1	&	vi	&	 P.Simecek 	;	 BRNO 23 	&	54058.5920	&	1	&	vi	&	 C.Stephan 	;	 AOEB 12 	\\
43725.5179	&	2	&	pe	&	 Z.Tufekcioglu 	;	IBVS 1495 	&	54059.3020	&	1	&	pe	&	 H.V. Senavci et al. 	;	IBVS 5754 	\\
43729.4333	&	1	&	pe	&	 Z.Tufekcioglu 	;	IBVS 1495 	&	54063.5720	&	1	&	vi	&	 C.Stephan 	;	 AOEB 12 	\\
43729.4380	&	1	&	vi	&	 P.Ivan 	;	 BRNO 23 	&	54070.3254	&	2	&	pe	&	 H.V. Senavci et al. 	;	IBVS 5754 	\\
43756.4831	&	1	&	pe	&	 Z.Tufekcioglu 	;	IBVS 1495 	&	54096.3177	&	1	&	CCD	&	 R.Ehrenberger 	;	OEJV 0074 	\\
43776.4140	&	1	&	vi	&	 D.Lichtenknecker 	;	BAVM 31 	&	54298.4676	&	1	&	vi	&	 M.Mruz 	;	OEJV 0094 	\\
43780.3277	&	2	&	pe	&	 Z.Tufekcioglu 	;	IBVS 1495 	&	54309.5089	&	2	&	CCD	&	 S.Parimucha et al. 	;	IBVS 5898 	\\
43791.3540	&	1	&	vi	&	 R.Germann 	;	 BBS 39 	&	54335.4878	&	1	&	pe	&	 T.Kilicoglu et al. 	;	IBVS 5801 	\\
43791.3700	&	1	&	vi	&	 H.Peter 	;	 BBS 39 	&	54335.4887	&	1	&	CCD	&	 L.melcer 	;	OEJV 0074 	\\
43802.7600	&	1	&	vi	&	 G.Samolyk 	;	 AOEB 2 	&	54345.4486	&	1	&	CCD	&	S.Caliskan	;	Nat. Ast. Cong., 2008	\\
43803.4650	&	1	&	vi	&	 H.Peter 	;	 BBS 39 	&	54351.5003	&	2	&	CCD	&	S.Caliskan	;	Nat. Ast. Cong., 2008	\\
43806.3090	&	1	&	vi	&	 R.Germann 	;	 BBS 39 	&	54394.5693	&	1	&	CCD	&	 G.Samolyk 	;	JAAVSO 36(2);171 	\\
43863.2560	&	1	&	vi	&	 R.Germann 	;	 BBS 41 	&	54416.6361	&	1	&	CCD	&	 J.Bialozynski 	;	JAAVSO 36(2);171 	\\
43878.2020	&	1	&	vi	&	 K.Locher 	;	 BBS 41 	&	54436.5670	&	1	&	CCD	&	 S.Dvorak 	;	IBVS 5814 	\\
44077.5070	&	1	&	vi	&	 D.Svelohva 	;	 BRNO 23 	&	54710.6180	&	1	&	CCD	&	 G.Samolyk 	;	JAAVSO 36(2);186 	\\
44092.4600	&	1	&	vi	&	 K.Locher 	;	 BBS 44 	&	54738.3787	&	1	&	CCD	&	 S.Parimucha et al. 	;	IBVS 5898 	\\
44102.4260	&	1	&	vi	&	 V.Wagner 	;	 BRNO 23 	&	54774.6840	&	1	&	CCD	&	 R.Diethelm 	;	IBVS 5871 	\\
44117.3690	&	1	&	vi	&	 R.Germann 	;	 BBS 44 	&	54799.5955	&	1	&	CCD	&	 K.Menzies 	;	JAAVSO 37(1);44 	\\
44117.3770	&	1	&	vi	&	 H.Peter 	;	 BBS 44 	&	55044.4620	&	1	&	CCD	&	 N.Erkan et al. 	;	IBVS 5924 	\\
44134.4580	&	1	&	vi	&	 H.Peter 	;	 BBS 45 	&	55064.3929	&	1	&	CCD	&	 G.-U.Flechsig 	;	BAVM 212 	\\
44143.3560	&	2	&	pe	&	 Z.Aslan et al. 	;	IBVS 1908 	&	55085.7474	&	1	&	CCD	&	 G.Samolyk 	;	 JAAVSO 38;120 	\\
44144.4227	&	1	&	pe	&	 Z.Aslan et al. 	;	IBVS 1908 	&	55116.3557	&	1	&	CCD	&	 N.Erkan et al. 	;	IBVS 5924 	\\
44164.3545	&	1	&	pe	&	 U.S.Chaubey 	;	 ASS 81.283 	&	55429.5569	&	1	&	CCD	&	 S.Dogru et al. 	;	IBVS 5988 	\\
44166.4920	&	1	&	vi	&	 T.Brelstaff 	;	 VSSC 59.19 	&	55498.2485	&	2	&	CCD	&	 S.Parimucha et al. 	;	IBVS 5980 	\\
44189.2670	&	1	&	vi	&	 H.Peter 	;	 BBS 45 	&	55524.9404	&	1	&	CCD	&	 K.Hirosawa 	;	VSB 51 	\\
44219.1650	&	1	&	pe	&	 U.S.Chaubey 	;	 ASS 81.283 	&	55561.2440	&	1	&	CCD	&	 L.melcer 	;	OEJV 0137 	\\
44435.5650	&	1	&	vi	&	 K.Locher 	;	 BBS 49 	&	55820.3460	&	1	&	CCD	&	 A.Paschke 	;	OEJV 0142 	\\
44440.5400	&	1	&	vi	&	 R.Germann 	;	 BBS 49 	&	55820.3461	&	1	&	CCD	&	 M.Dietrich 	;	BAVM 225 	\\
44445.5250	&	1	&	vi	&	 K.Locher 	;	 BBS 49 	&	55867.3270	&	1	&	CCD	&	 L.melcer 	;	OEJV 0160 	\\
44455.4900	&	1	&	vi	&	 K.Chyzny 	;	 MVS 9.18 	&	55887.2592	&	1	&	CCD	&	 D.Buhme 	;	BAVM 225 	\\
44470.4380	&	1	&	vi	&	 P.Kucera 	;	 BRNO 23 	&	56163.4447	&	1	&	CCD	&	 S.Parimucha et al. 	;	IBVS 6044 	\\
44474.7030	&	1	&	vi	&	 G.Samolyk 	;	 AOEB 2 	&	56210.0691	&	2	&	CCD	&	Y. Yang	;	AJ 147	\\
44490.3640	&	1	&	vi	&	 R.Diethelm 	;	 BBS 50 	&	56211.1365	&	1	&	CCD	&	Y. Yang	;	AJ 147	\\
44490.3660	&	1	&	vi	&	 H.Peter 	;	 BBS 50 	&	56212.2052	&	2	&	CCD	&	Y. Yang	;	AJ 147	\\
44497.4860	&	1	&	vi	&	 G.Mavrofridis 	;	 BBS 51 	&	56219.6785	&	1	&	CCD	&	 G.Frey 	;	 JAAVSO 42;426 	\\
44502.4654	&	1	&	pe	&	 D.Elias 	;	 BBS 54 	&	56229.6439 	&	1	&	CCD	&	 G.Frey 	;	 JAAVSO 42;426 	\\
44502.4690	&	1	&	vi	&	 D.Mourikis 	;	 BBS 50 	&	56231.7796 	&	1	&	CCD	&	 G.Frey 	;	 JAAVSO 42;426 	\\
44512.4340	&	1	&	vi	&	 H.Peter 	;	 BBS 50 	&	56256.6934 	&	1	&	CCD	&	 G.Frey 	;	 JAAVSO 42;426 	\\
44517.4160	&	1	&	vi	&	 G.Mavrofridis 	;	 BBS 51 	&	56501.5600	&	1	&	CCD	&	  K.Rutz 	;	BAVM 234 	\\
44517.4190	&	1	&	vi	&	 G.Stefanopoulos 	;	 BBS 52 	&	56537.8635	&	1	&	CCD	&	 G.Samolyk 	;	 JAAVSO 41;328 	\\
44524.5340	&	1	&	vi	&	 G.Mavrofridis 	;	 BBS 51 	&	56557.7934	&	1	&	CCD	&	 B.Manske 	;	 JAAVSO 41;328 	\\
44532.3640	&	1	&	vi	&	 W.Braune 	;	BAVM 32 	&	56557.7946	&	1	&	CCD	&	 G.Frey 	;	 JAAVSO 42;426 	\\
44543.0401	&	1	&	pe	&	 H.D.Kennedy 	;	IBVS 2118 	&	56565.6246	&	1	&	CCD	&	 B.Manske 	;	 JAAVSO 41;328 	\\
44557.9879	&	1	&	pe	&	 H.D.Kennedy 	;	IBVS 2118 	&	56567.7599	&	1	&	CCD	&	 G.Frey 	;	 JAAVSO 42;426 	\\
44567.2420	&	1	&	vi	&	 H.Peter 	;	 BBS 51 	&	56572.7430	&	1	&	CCD	&	 G.Frey 	;	 JAAVSO 42;426 	\\
44567.2450	&	1	&	vi	&	 R.Germann 	;	 BBS 51 	&	56577.7255 	&	1	&	CCD	&	 G.Frey 	;	 JAAVSO 42;426 	\\
44593.5840	&	1	&	vi	&	 G.Hanson 	;	 AOEB 2 	&	56587.6911 	&	1	&	CCD	&	 G.Frey 	;	 JAAVSO 42;426 	\\
44636.2870	&	1	&	vi	&	 R.Germann 	;	 BBS 52 	&	56588.4035 	&	1	&	CCD	&	 F.Agerer 	;	BAVM 234 	\\
44823.4940	&	1	&	vi	&	 T.Kaczkowski 	;	 MVS 9.90 	&	56597.6568	&	1	&	CCD	&	 G.Frey 	;	 JAAVSO 42;426 	\\
44823.5000	&	1	&	vi	&	 T.Graf 	;	 BRNO 26 	&	56602.6394 	&	1	&	CCD	&	 G.Frey 	;	 JAAVSO 42;426 	\\
44843.4272	&	1	&	pe	&	 E.Derman et al. 	;	IBVS 2159 	&	56905.5192	&	2	&	CCD	&	M. Masek 	;	BRNO 40	\\
44848.4102	&	1	&	pe	&	 E.Derman et al. 	;	IBVS 2159 	&	56929.3667	&	1	&	CCD	&	 F.Agerer 	;	BAVM 239 	\\
44853.3920	&	1	&	vi	&	 H.Peter 	;	 BBS 57 	&	56930.4362	&	1	&	CCD	&	 F.Agerer 	;	BAVM 239 	\\
44853.3950	&	1	&	vi	&	 K.Carbol 	;	 BRNO 26 	&	56953.5685	&	1	&	CCD	&	 N.Simmons 	;	 JAAVSO 43-1 	\\
44883.2830	&	1	&	vi	&	 N.Stoikidis 	;	 BBS 57 	&	56955.7049	&	1	&	CCD	&	 G.Frey 	;	 JAAVSO 44-1 	\\
44890.4100	&	1	&	vi	&	 H.Peter 	;	 BBS 57 	&	57251.8222	&	1	&	CCD	&	 K.Menzies 	;	 JAAVSO 43-2 	\\
44893.2550	&	1	&	vi	&	 N.Stoikidis 	;	 BBS 57 	&	57267.4823	&	1	&	CCD	&	E. Bahar	;	IBVS 6209	\\
44900.3870	&	1	&	vi	&	 G.Mavrofridis 	;	 BBS 57 	&	57278.5163	&	1	&	CCD	&	 F.Agerer 	;	IBVS 6196	\\
44910.3300	&	1	&	vi	&	 N.Stoikidis 	;	 BBS 57 	&	57308.7680	&	1	&	CCD	&	 G.Frey 	;	 JAAVSO 44-1 	\\
44925.2840	&	1	&	vi	&	 H.Peter 	;	 BBS 57 	&	57327.2750 	&	2	&	CCD	&	 S.Parimucha	;	IBVS 6167 	\\
45170.8580	&	1	&	vi	&	 E.Halbach 	;	 AOEB 2 	&	57390.6267 	&	1	&	CCD	&	 R.Sabo 	;	 JAAVSO 44-1	\\
45196.4870	&	1	&	pe	&	 A.Buchtler 	;	IBVS 2385 	&	58059.3779	&	2	&	CCD	&	our study	;	--	\\
45201.4690 	&	1	&	pe	&	 M.Prikryl 	;	 BRNO 26 	&	58060.4456	&	1	&	CCD	&	our study	;	--	\\
\hline
\end{longtable}
  \endgroup

Thus, the new ephemeris was calculated as;
\begin{equation}
HJD_{\mathrm{Min I}} = 2455867.327300(81) + 0_{\cdot}^\mathrm{d}711816455(19)\times E.
\end{equation}

The {\sl O-C} diagram shown in Figure~1 (top panel) displayed two sinusoidal curves superimposed on each other. Of which, the primary curve had an eccentric cyclic change which had almost three maximum and two minima. Also, the residuals from the sinusoidal fit showed another low-amplitude, short-period and eccentric cyclic modulation having three minima and four maxima. Our observational CCD minima were the last two points plotted on the {\sl O-C} diagram. These points allowed us to determine the turn point of the last maximum of the {\sl O-C} curve.   

We first used the PERIOD04 program (Lenz and Breger 2005) to analyze the weighted data. Then, we extracted the individual frequencies causing the fluctuations. Two frequencies of $f_1 = 0.000041375$ c/E ($A_1$=0.0082, S/N = 7.84) and $f_2$ = 0.000072382 c/E ($A_1$=0.0059, S/N = 18.04), shown in Figure~2, were detected. These frequencies corresponded to two periods of $47.10\pm0.63$ and $26.92\pm0.44$ years, respectively. When these two theoretical frequencies  were adjusted to the {\sl O-C} diagram in Figure~1, they were in good agreement with observational data. For the eccentricities seen in the curves, the light-time effect caused by the third and fourth bodies in the system was considered. In order to derive light-time orbits and the parameters of the third and fourth additional bodies, we used the equations given by Irwin (1952). Furthermore, the computer code called OC2LTE30 (Ak \etal 2004) was used to determine the orbital parameters. All of these results are presented in Table~3.

In Figure~1, the orbital parameters of the third and fourth body are presented in the second and the third panels. The sum of these lines, which corresponds to the total theoretical {\sl O-C} curve, are shown as the continuous line in the first panel. The sum of the least squares of the total residuals is $1.6\times 10^{-2}$ $day^{2}$. The estimated errors of these parameters arise from the non-linear least-squares method, on which the inverse problem solving method is based. This method does not take into account the error of each observation point and the possible correlations of fitted parameters with each other. Therefore, the standard error values given for the parameters may be smaller than they should be. So, the standard error values given in the table should be considered as the lowest limits. 

\IBVSfig{20cm}{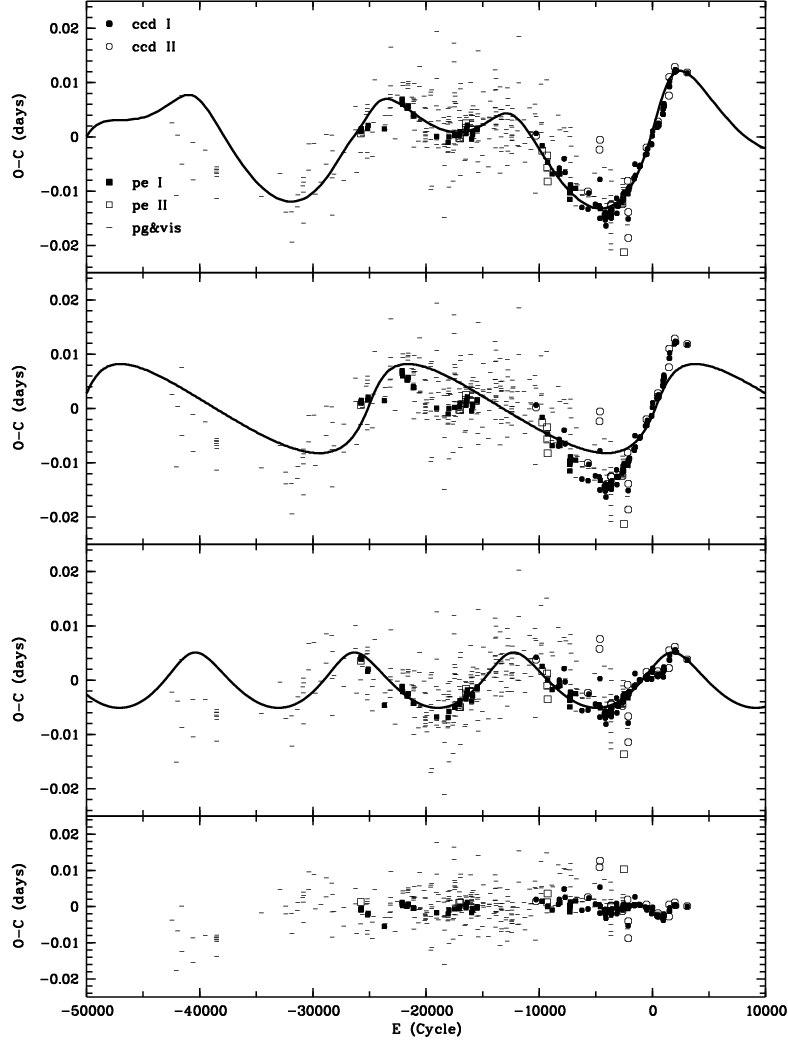}{The O-C diagram of DI Peg. The first panel shows the overall data and the total theoretical O-C variation (continuous line). While the second panel presents the primary and highly eccentric sinosoidal variation, the residual data which have another sinusoidal modulation are displayed in the third panel. The final residuals are given in the last panel.}  
\IBVSfigKey{f1.jpg}{DI Peg}{O-C diagram}

\IBVSfig{10cm}{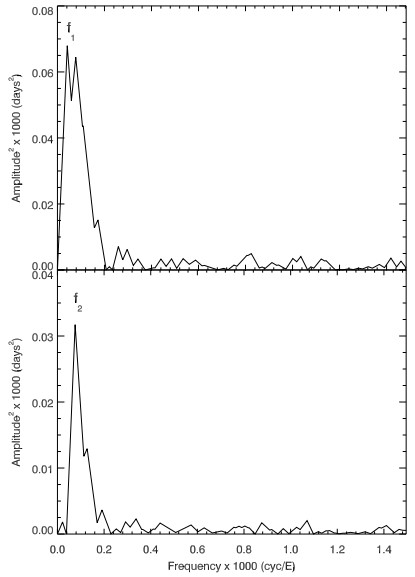}{The two frequencies of $f_1$ = 0.000041375 and $f_2$ = 0.000072382 c/E  detected by PERIOD04.}
\IBVSfigKey{f2.jpg}{DI Peg}{amplitude diagram}

\vskip 1cm

\leftline{Table 3. Parameters and standard errors derived from {\sl O-C} analysis of each additional body.}
\vskip 3mm
\begin{center}
\begin{tabular}{l @{~~~~}c c @{}c c @{}c }\hline
Parameters & Third Body & & Fourth Body \\
\hline
{\it P}$_{3,4}$ [years]                      	   &	$ 49.50\pm0.36$ &	& $27.40\pm0.24$&	\\
{\it A} [days]                              		  &	 $0.0082\pm0.0002$ & & $0.0051\pm0.0002$ &\\

${\it e}^{'}$             &	$0.61\pm0.06$ &	&$0.30\pm0.08$ &	\\
$\omega ^{'}$ [$^\circ$]    &	$7.00\pm1.74$	&	&$75.00\pm3.63$ &	\\
{\it T}$^{'}$ [HJD]       	&$2456220\pm 261$	&	&$2456860\pm150$ &	\\

{\it f}~({\it $m_{3,4}$}) [$M_\odot$]             &	$0.0023 \pm 0.0007$	& &	$0.0009 \pm 0.0001$\\
{\it m} [M$_\odot$]          	 	 &	$0.2135 \pm 0.0213$	& &	$0.1505 \pm 0.0075$\\
{\it L}$_{Bol}$ [L$_\odot$]           	 &	$0.0061 \pm 0.0017$	& &	$0.0025 \pm 0.0003$\\
{\it M}$_{Bol}$ [mag]           	 &	$10.23 \pm 0.27$	& &	$11.22 \pm 0.14$\\
{\it m}$_{Bol}$ [mag]           	 &	$18.22 \pm 1.38$	& &	$19.21 \pm 1.24$\\
$\theta$ [arcsec]           	 &	$0.0915 \pm 0.0277$	& &	$0.0625 \pm 0.0184$\\
{$\sum(O-C)^2$} [$day^{2}$] & $260\times10^{-4}$ & & $138\times10^{-4}$ \\

\hline
\end{tabular}
\end{center}
\vskip 1cm

\section{Results and Discussion}
\label{conclusion}

An {\sl O-C} diagram is a special plot generally used to determine period changes that are difficult to detect by direct measurements. If there is not any measurable change in period, then the {\sl O-C} difference generates a straight line. If any variation in period is detected, however, the {\sl O-C} data generate a structure that displays the characteristic of the mechanism causing this variation. These mechanisms can be arranged as: mass transfer between components or mass loss from the system, spin-orbital interactions, angular momentum loss through stellar winds, gravitational waves, oscillations in rotation, differential rotation, apsidal motion, presence of a third light, and magnetic activity (Mikulasek \etal 2012).   

In terms of binarity, orbital period change is quite an important subject since it is related to the formation, structure, and evolution of binary stars. These variables gain and lose mass and angular momentum as specified by Roche geometry. These events are the first proposed mechanisms to explain observed period changes. Both of these mechanisms can increase or decrease the period of the system and generate parabolic structures in the {\sl O-C} diagram. The mass transfer between components is more effective in changing the orbital period than the mass loss from the system. The most basic case to be considered for exchanging material between components is conservative mass transfer. In this case, the mass lost by one component is gained by the companion star, so the total mass of the system and thus the total orbital angular momentum is preserved.       

Among the common mechanisms given above, apsidal motion involves a change in the orientation of the system's major axis, since the potential energy between the components does not exactly obey Newton's gravitational law. In the {\sl O-C} diagram, the times for secondary and primary minima shift in opposite directions. However, as this mechanism requires large eccentricities, it is rarely observed (Zavala \etal 2002). Alternatively, it is assumed that the cyclic pattern is caused by the presence of a third body in the system. Based on this assumption, the primary and secondary eclipse times are produced by the motion of the binary around the common centre of mass of a triple system. In this case, the periodic pattern arises from the light-time effect (Borkovits and Hegedues 1996). 

Apart from these, another mechanism to cause period variation in binary stars is magnetic activity cycles. In the systems having late-type components, if the shape of the companion star is distorted by tidal and centrifugal forces, changes in the internal rotation associated with a magnetic activity cycle vary the gravitational quadrupole moment. As the quadrupole moment increases, the gravitational field increases leading to a decrease in the period. Otherwise, if the quadrupole moment decreases, the orbital period increases (Applegate 1992). Magnetic activity produces cyclic modulations in the {\sl O-C} diagram, and their periods are from years to decades. 

In Algols, alternate orbital period changes are well known in systems with a late-type secondary star (Zavala \etal 2002). For a binary system, cyclic period variability are generally thought to be caused by either magnetic activity in one or both components (Applegate 1992) or light-time effect due to a third body (Irwin 1952). In terms of magnetic activity, observed oscillations are arisen from the variations of the gravitational quadrupole moment ($\Delta Q$), which is typically around $10^{51}-10^{52}$ g cm$^2$ for close binaries and can be calculated from the equation of

\begin{equation}
\frac{\Delta P}{P} = \frac{-9\Delta Q}{M a^2} \approx \frac{2\pi A_{\mathrm{sin}}}{P_{\mathrm{sin}}}
\end{equation}

\noindent where $M$ is the mass of the active component (Lanza 2002). 

In the case of DI\,Peg, the {\sl O-C} diagram shows neither a parabolic change which is an indication of a mass transfer between the components or a mass loss from the system, nor anti-correlation between the primary and secondary minimum timings that is a sign for a change in the orientation of the binary's major axis. On the other hand, it is known that the star has a late-type companion (K4). For this reason, there is a potential that this component may show magnetic activity. In order to search this possibility, we calculate the gravitational quadrupole moment ($\Delta Q$) of the secondary star by using $\Delta P/P = 3.20 \times 10^{-6}$ which is calculated in this study and by adopting $M_1$ = 1.18(3) M$_\odot$, $M_2$ = 0.70(2) M$_\odot$, and $a$ = 4.14(5) R$_\odot$ from Lu (1992). As a result, we find the variation of the quadrupole moment of the star to be $\Delta Q_2 = 4.11 \times 10^{49}$ g cm$^2$. Since this result is clearly smaller than the typical value and the sinusoidal variations are eccentric, it is unlikely that magnetic activity is responsible for the periodic modulations in DI\,Peg. 

Therefore, two sinusoidal changes can be more likely attributed to the light-time effects due to the presence of two additional bodies. Since the third body is confirmed from the spectroscopic study by Lu (1992), we calculate the specific parameters of the third body under the assumption of the presence of an object gravitationally bound to the system. From the {\sl O-C} diagram, the period and amplitude of the primary modulation are found to be 49.50$\pm0.36$ yr and 0.0082 days. The projected distance of the mass center of the eclipsing pair to the center of mass of the triple system is around 1.78$\pm0.16$  au. By using these values the mass function of the third-body is found to be 0.0023(7).  If the third-body orbit is co-planar with the orbit of the system (i.e., $i \sim 90^{\circ}$), its mass would be 0.21(2) M$_{\odot}$. Also, from the Kepler's third law, the semi-major axis of the orbit is computed as 15.75(7) au. By adopting the parallax of the star from van Leeuwen (2007), we derive the distance of d $\sim 191(43)$ parsecs and hence the maximum angular separation of the third-body from the eclipsing pair to be 0.091(28) arcsec. Using the mass-luminosity relation for main-sequence stars given by Demircan and Kahraman (1991), we can estimate the bolometric absolute magnitude of the third body for the given distance to be about $M_{bol}$ = 10.23(27) mag. According to Allen's table (Cox 2000), the spectral type for the third body can be estimated to be M3, which points a red dwarf. 

Additionally, as mentioned in the previous section, the residuals from the sine fit show another low-amplitude, short-period and eccentric cyclic modulation. This variation is also interpreted as the existence of a fourth body physically connected to the system by Yang \etal (2014). From the {\sl O-C} diagram, we calculated the period and amplitude of the secondary modulation as 27.40(24) yr and  0.0051(2) days. The mass function and the mass of the fourth body are $f(m_4)$ = 0.0009(1) and $M_4$ = 0.151(75) M$_\odot$. Assuming that the object orbits in the same plane as the system and taking the aforementioned distance value into account, we find the angular separation of the fourth body from the eclipsing pair to be 0.0615(183) arcsec. By using the mass-luminosity relation for main-sequence stars given by Demircan and Kahraman (1991), we estimate the bolometric absolute magnitude of the fourth body to be about $M_{bol}$ = 11.22(14) mag. According to Allen's table (Cox 2000), the additional fourth body may be a M4 spectral type red dwarf.  

Additionally, from Figure~1, the residuals of two sinusoidal fits still seem to show another modulation. The period and amplitude of this modulation are roughly $P = 19.5$ years and $A = 0.004$ days. However, it is not possible to attribute this change as another object that is in orbit around the binary system. Therefore, we recommend future photometric and spectroscopic observations to reveal the true nature of DI Peg.

\bigskip
{\bf \noindent Acknowledgements}
We thank Ankara \"{U}niversity Kreiken Observatory for the support of project number T35\_2017\_IV\_06. This research has made use of the SIMBAD database, operated at CDS, Strasbourg, France, and of NASA's Astrophysics Data System Bibliographic Services.

\vskip 1cm

\references

Ahnert P., 1974, {\it MitVS}, {\bf 6}, 158 \BIBCODE{1974MitVS...6..158A}

Ak, T., Albayrak, B., Selam, S.O., Tanriverdi, T.: 2004, {\it NewA}, {\bf 9}, 265 \BIBCODE{2004NewA....9..265A}

Applegate, J.H., 1992, {\it Astrophys. J.} {\bf 385}, 621 \BIBCODE{1992ApJ...385..621A}

Binnendijk, L. 1973, {\it AJ}, {\bf 78}, 97   \BIBCODE{1973AJ.....78...97B}

Borkovits, T.~and Hegedues, T., 1996, {\it Astronomy and Astrophysics Supplement Series} {\bf 120}, 63 \BIBCODE{1996A&AS..120...63B} 

Borkovits, T., Rappaport, S., Hajdu, T., and Sztakovics, J., 2015, {\it Monthly Notices of the Royal Astronomical Society} {\bf 448}, 946 \BIBCODE{2015MNRAS.448..946B}

Borkovits, T., Hajdu, T., Sztakovics, J., Rappaport, S., Levine, A., B{\'{\i}}r{\'o}, I.B., and Klagyivik, P., 2016, {\it Monthly Notices of the Royal Astronomical Society} {\bf 455}, 4136 \BIBCODE{2016MNRAS.455.4136B}

Chaubey, U. S., 1982, {\it Ap\&SS}, {\bf 81}, 283  \BIBCODE{1982Ap&SS..81..283C}

Chauvin G., Lagrange A.-M., Udry S., Mayor M., 2007, {\it A\&A}, {\bf 475}, 723  \BIBCODE{2007A&A...475..723C}

Chou, K. C., ~and Kitamura, M., 1968, {\it JKAS}, {\bf 1}, 1 \BIBCODE{1968JKAS....1....1C}

Conroy, K.E., Pr{\v s}a, A., Stassun, K.G., Orosz, J.A., Fabrycky, D.C., and Welsh, W.F., 2014, {\it The Astronomical Journal} {\bf 147}, 45 \BIBCODE{2014AJ....147...45C}

Cox A. N., 2000, {\it asqu.book}  \BIBCODE{2000asqu.book.....C}

Demircan O., Kahraman G., 1991, {\it Ap\&SS}, {\bf 181}, 313 \BIBCODE{1991Ap&SS.181..313D}

Dvorak R., 1986, {\it A\&A}, {\bf 167}, 379 \BIBCODE{1986A&A...167..379D}

Evans D. S., 1968, {\it QJRAS}, {\bf 9}, 388  \BIBCODE{1968QJRAS...9..388E}

Furlan E., et al., 2007, {\it ApJ}, {\bf 664}, 1176  \BIBCODE{2007ApJ...664.1176F}

Hanna M. A., Amin S. M., 2013, {\it JKAS}, {\bf 46}, 151 \BIBCODE{2013JKAS...46..151H}

Gaposchkin, S., 1953, {\it AnHar}, {\bf 113}, 67C  \BIBCODE{1953AnHar.113...67G}

Irwin J. B., 1952, {\it ApJ}, {\bf 116}, 211  \BIBCODE{1952ApJ...116..211I}

Jensch, A., 1934, {\it AN}, {\bf 252}, 395  \BIBCODE{1934AN....252..393J}

Kruszewski, A.: 1964, {\it AcA}, {\bf 14}, 241   \BIBCODE{1964AcA....14..241K}
 
Kwee K. K., van Woerden H., 1956, {\it BAN}, {\bf 12}, 327  \BIBCODE{1956BAN....12..327K}

Lanza, A.F.~and Rodon{\`o}, M.: 2002, {\it Astronomische Nachrichten} {\bf 323}, 424 \BIBCODE{2002AN....323..424L}

Lee J. W., Lee C.-U., Kim S.-L., Kim H.-I., Park J.-H., 2012, {\it AJ}, {\bf 143}, 34   \BIBCODE{2012AJ....143...34L}

Lehmann, H., Borkovits, T., Rappaport, S.A., Ngo, H., Mawet, D., Csizmadia, S., and Forg{\'a}cs-Dajka, E., 2016, {\it Astrophys. J.} {\bf 819}, 33 \BIBCODE{2016ApJ...819...33L}

Lim J., Takakuwa S., 2006, {\it ApJ}, 653, 425  \BIBCODE{2006ApJ...653..425L}

Lenz, P., ~and Breger, M.: 2005, {\it Commun. Asteroseismol.}, {\bf 146}, 53 \BIBCODE{2005CoAst.146...53L}

Lu, W., 1992, {\it AcA}, {\bf 42}, 77C   \BIBCODE{1992AcA....42...73L}

Marzari F., Scholl H., Thébault P., Baruteau C., 2009, {\it A\&A}, {\bf 508}, 1493  \BIBCODE{2009A&A...508.1493M}

Mikul{\'a}{\v s}ek, Z., Zejda, M., and Jan{\'{\i}}k, J., 2012, {\it From Interacting Binaries to Exoplanets: Essential Modeling Tools} {\bf 282}, 391 \BIBCODE{2012IAUS..282..391M}

Morgenroth, O., 1934, {\it AN}, {\bf 252}, 389  \BIBCODE{1934AN....252..389M}

Neuhäuser R., Mugrauer M., Fukagawa M., Torres G., Schmidt T., 2007, {\it A\&A}, {\bf 462}, 777  \BIBCODE{2007A&A...462..777N}

Pribulla T., et al., 2012, {\it AN}, {\bf 333}, 754  \BIBCODE{2012AN....333..754P}

Qian, S., 2001, {\it MNRAS}, {\bf 328}, 914 \BIBCODE{2001MNRAS.328..914Q} 

Rafert J. B., 1982, {\it PASP}, {\bf 94}, 485  \BIBCODE{1982PASP...94..485R}

Rappaport, S., Lehmann, H., Kalomeni, B., Borkovits, T., Latham, D., Bieryla, A., Ngo, H., Mawet, D., Howell, S., Horch, E., Jacobs, T.L., LaCourse, D., S{\'o}dor, {\'A}., Vanderburg, A., and Pavlovski, K., 2016, {\it Monthly Notices of the Royal Astronomical Society} {\bf 462}, 1812 \BIBCODE{2016MNRAS.462.1812R}

Rappaport, S., Vanderburg, A., Borkovits, T., Kalomeni, B., Halpern, J.P., Ngo, H., Mace, G.N., Fulton, B.J., Howard, A.W., Isaacson, H., Petigura, E.A., Mawet, D., Kristiansen, M.H., Jacobs, T.L., LaCourse, D., Bieryla, A., Forg{\'a}cs-Dajka, E., and Nelson, L., 2017, {\it Monthly Notices of the Royal Astronomical Society} {\bf 467}, 2160 \BIBCODE{2017MNRAS.467.2160R}

Reipurth B., 2000, {\it AJ}, {\bf 120}, 3177  \BIBCODE{2000AJ....120.3177R}	

Rucinski, R., 1967, {\it AcA}, {\bf 17}, 271  \BIBCODE{1967AcA....17..271R}	

Schwarz R., Haghighipour N., Eggl S., Pilat-Lohinger E., Funk B., 2011, {\it MNRAS}, {\bf 414}, 2763  \BIBCODE{2011MNRAS.414.2763}

van den Berk J., Portegies Zwart S. F., McMillan S. L. W., 2007, {\it MNRAS}, {\bf 379}, 111  \BIBCODE{2007MNRAS.379..111V}

van Leeuwen F., 2007, {\it A\&A}, {\bf 474}, 653   \BIBCODE{2007A&A...474..653V}	

Vinko, J., 1992, {\it IBVS}, {\bf 3757}, 1   \BIBCODE{1992IBVS.3757....1V}

Yang Y.-G., Yang Y., Li S.-Z., 2014, {\it AJ}, {\bf 147}, 145   \BIBCODE{2014AJ....147..145Y}

Zavala, R.T., McNamara, B.J., Harrison, T.E., Galvan, E., Galvan, J., Jarvis, T., Killgore, G., Mireles, O.R., Olivares, D., Rodriguez, B.A., Sanchez, M., Silva, A.L., Silva, A.L., and Silva-Velarde, E., 2002, {\it The Astronomical Journal} {\bf 123}, 450. \BIBCODE{2002AJ....123..450Z}

\endreferences

\end{document}